\documentclass[manuscript,twoside]{aastex}

\newcommand{\Vtworev}{\marginpar{\centerline{\scriptsize rev}}}
\renewcommand{\Vtworev}{\relax}

\usepackage{xspace}
\usepackage{amssymb}

\newcommand{\tooe}  { {$\theta^1\,\mathrm{Ori~E}$}\xspace}
\newcommand{\tooc}  { {$\theta^1\,\mathrm{Ori~C}$}\xspace}
\newcommand{\torie} {{$\theta^1$ Ori E}\xspace}

\newcommand{\toriaa}{{$\theta^2$ Ori A}\xspace}

\newcommand{\mone}  {^{-1}}
\newcommand{\mtwo}  {^{-2}}
\newcommand{\mthree}{^{-3}}
\newcommand{\mfour} {^{-4}}

\newcounter{ion}
  \newcommand{\eli}[2]
  {\setcounter{ion}{#2}#1{~\sc\roman{ion}}}

\newcommand{\aped}  {{APED}\xspace}

\newcommand{\chan}  {{\it Chandra}\xspace}

\newcommand{\ciao}  {{CIAO}\xspace}

\newcommand{\heg}   {{HEG}\xspace}

\newcommand{\hetg}  {{HETG}\xspace}
\newcommand{\hetgs} {{HETGS}\xspace}

\newcommand{\meg}   {{MEG}\xspace}
\newcommand{\neix}  {\eli{Ne}{9}}
\newcommand{\mgxi}  {\eli{Mg}{11}}
\newcommand{\ovii}  {\eli{O}{7}}

\newcommand{\ang} {{\AA}\xspace}

\def\chandra{{\sl Chandra}}

\def\ergsec{\mathrm{ergs\, s^{-1}}} 

\def\Msun{$\mathrm{M_{\odot}}$ } 
\def\Rsun{$\mathrm{R_{\odot}}$ }
\def\lsun{$\mathrm{L_{\odot}}$ }

\def\it{\sl}

\def\th1e{$\theta^1$ Ori E}
\def\th1c{$\theta^1$ Ori C}
\def\th2a{$\theta^2$ Ori A}
\newcommand{\msun}{\mathrm{M_{\odot}}\xspace} 
\newcommand{\rsun}{\mathrm{R_{\odot}}}
\newcommand{\Lsun}{\mathrm{L_{\odot}}}

\newcommand{\note}[1]{\marginpaxor{\raggedright\scriptsize #1}} 
\renewcommand{\note}[1]{\relax}

\shorttitle{X-ray Emission and Corona of $\theta^1$ Ori E}
\shortauthors{Huenemoerder et al.}

\begin{document}



\title{X-ray Emission and Corona of the Young Intermediate Mass Binary
  $\theta^1$ Ori E}

\author{David P. Huenemoerder\altaffilmark{1},\\
Norbert S. Schulz\altaffilmark{1},\\
Paola Testa\altaffilmark{2},\\
Anthony Kesich\altaffilmark{3},\\
\& Claude R. Canizares\altaffilmark{1}\\
\ \\
}

\altaffiltext{1}{Massachusetts Institute of Technology,
  Kavli Institute for Astrophysics and Space 
  Research, 70 Vassar 
  St., Cambridge, MA, 02139}   

\altaffiltext{2}{Harvard-Smithsonian Center for Astrophysics,
  60 Garden St., Cambridge, MA, 02138}   

\altaffiltext{3}{University of California - Davis, One Shields Ave.,
  Davis, CA 95616 }


\begin{abstract}
  \torie is a young, moderate mass binary system, a rarely observed
  case of spectral-type G-giants of about 3 Solar masses, which are
  still collapsing towards the main sequence, where they presumably
  become X-ray faint.  We have obtained high resolution X-ray spectra
  with \chandra\ and find that the system is very active and similar
  to coronal sources, having emission typical of magnetically confined
  plasma: a broad temperature distribution with a hot component and
  significant high energy continuum; narrow emission lines from H- and
  He-like ions, as well as a range of Fe ions, and relative
  luminosity, $L_x/L_\mathrm{bol} = 10\mthree$, at the saturation
  limit.  Density, while poorly constrained, is consistent with the
  low density limits, our upper limits being
  $n_e<10^{13}\,\mathrm{cm\mthree}$ for \eli{Mg}{11} and
  $n_e<10^{12}\,\mathrm{cm\mthree}$ for \eli{Ne}{9}.  Coronal
  elemental abundances are sub-Solar, with Ne being the highest at
  about 0.4 times Solar. We find a possible trend in Trapezium hot
  plasmas towards low relative abundances of Fe, O, and Ne, which is
  hard to explain in terms of the dust depletion scenarios of low-mass
  young stars.  Variability was unusually low during our observations
  relative to other coronally active stars.  Qualitatively, the
  emission is similar to post main-sequence G-stars.  Coronal
  structures could be compact, or comparable to the dimensions of the
  stellar radii.  From comparison to X-ray emission from similar mass
  stars at various evolutionary epochs, we conclude that the X-rays in
  \tooe are generated by a convective dynamo, present during
  contraction, but which will vanish during the main-sequence epoch,
  and possibly to be resurrected during post main-sequence evolution.
\end{abstract}


\keywords{stars: individual (tet01 Ori E);  stars: coronae; stars:
  pre-main sequence; X-rays: stars}


\section{Introduction\label{sec:intro}}
\note{abstract updated} 

Intermediate mass pre-main sequence (PMS) stars, like their massive
cousins, are difficult to study because of their rapid evolutionary
time scales. Though not as short as stars above $8\,\msun$ which take
less than $10^5$ years to reach the main sequence, intermediate
mass stars between 2$\msun$ and $8\msun$ may only take 10--20 Myr.  In
both cases the accretion time scales dominate the evolution time to
the zero-age main sequence (ZAMS), in contrast to the low-mass T Tauri
stars for which PMS contraction times are longest. Intermediate mass
PMS stars are also not easily found and identified, and most existing
studies focus on Herbig Ae and Be (HAeBe) stars.  Herbig stars
\citep{Herbig:1960} are recognized as such once they already
contracted to high enough photospheric temperatures to be optically
identified as A and B stars and are thus already close to the ZAMS.
\Vtworev 
Herbig stars mark the transition between formation mechanisms of
low-mass and high-mass stars~\citep{Baines06}.  Herbig Ae stars seem
more similar to the low-mass T Tauri stars~\citep{Waters98, Vink05}.
Herbig Be stars are more similar to embedded young massive stars
\citep{Drew97}. Both the Ae and Be stars are all already in fairly
late PMS stages.

Most studies of Herbig stars use infrared and optical wavelengths to
probe their circumstellar disks and dusty environments.  Recent
studies suggest that specifically in HAe stars there is evidence not
only for circumstellar disks~\citep{Mannings97, Grady99} but also
indications of dust shadowing and settling indicative of dust grain
growth and planetesimal formation \citep{Acke04, Dullemond04,
  Grady05}. Some studies also suggest that magnetospheric accretion
analogous to classical T Tauri stars is possible ~\citep{Muzerolle04,
  Grady04, Guimaraes06}.  Recent modeling of 37 Herbig Ae/Fe stars
using UV spectra revealed that all but one show indications
of accretion with accretion rates in many cases substantially
exceeding $10^{-8}\, \msun \mathrm{yr^{-1}}$ \citep{Blondel06}.

Binarity also seems to be an important attribute in the formation and
evolution of intermediate mass stars.  In a sample of 28 HAeBe stars,
\citet{Baines06} find a binarity fraction of almost 70\% with a higher
binary frequency in HBe stars than in HAe stars.  HAe stars with close
companions also seem to lack circumstellar disks~\citep{Grady05}.

X-ray studies of young intermediate mass stars are still quite rare
and to date also focus almost entirely on HAe stars. Systematic
studies have shown that these are moderately bright in X-rays
~\citep{Damiani94, Zinnecker94, Hamaguchi05, Stelzer:al:2006}.  This fact is
already quite remarkable since main sequence A-stars lack strong winds
or coronae and it suggests that the physical characteristics of HAe
stars stars differ from those of main sequence A- and B-stars.
Mechanisms suggested range from active accretion to coronal activity
to some other form of plasma confinement.  
\Vtworev 
It is also possible, given the high frequency of binaries among HAeBe
stars, that some X-ray sources could be due to late-type companions
\citep{Stelzer:Hu:al:2006, Stelzer:al:2006}.
A detailed summary can be found in a recent \chandra\ high resolution
spectroscopic study of the HAe star HD~104237 in the $\epsilon$
Chamaeleontis Group \citep{Testa08}.

\Vtworev 
In this paper we focus on X-ray emission from \tooe, which was
recently determined to be an intermediate mass binary star.  The Orion
Trapezium is generally known for its ensemble of the nearest and
youngest massive stars
\citep{Schulz:Canizares:al:2001,Schulz:2003,Stelzer:al:2005}. Recent
studies now suggest the presence of several intermediate mass stars.
\toriaa\ harbors the second most massive O-star of the Trapezium, but
also two unidentified intermediate mass stars both between $3\msun$
and $7\msun$ \citep{Preibisch99}. The system is particularly
interesting in X-rays for its high luminosity and hard spectral
properties as well as giant hard X-ray outbursts \citep{Feigelson02,
  Schulz06}.  Plasma temperatures during these outbursts exceed $10^8
\mathrm{K}$ \citep{Schulz06}.  While the latter authors suggested a
possible link of these outbursts to binary interactions involving the
closer intermediate mass companion, there is also some evidence that
these may be connected to the more distant companion (M.\ Gagne,
private communication).

\torie\ is another system now known to contain young intermediate mass
stars.  It was long misidentified as B5 to B8 spectral type
\citep{Parenago54, Herbig:1960}. \citet{Herbig:Griffin:2006} obtained
optical spectroscopic radial velocity measurements and identified the
system as a binary containing two G~III type stars of masses of about
3--4 \Msun in a 9.9 day orbit.  Evolutionary tracks constrain the age
of the system to 0.5--1.0 Myr making the components of \torie\ some of
the youngest intermediate mass PMS stars known and far younger than
stars in the HAeBe phase. \torie\ is not among the optically brightest
stars in the Trapezium, but has long been recognized as the second
\Vtworev 
brightest Trapezium source in X-rays \citep{Ku82, Gagne94,
  Schulz:Canizares:al:2001}.  The \chandra\ Orion Ultradeep Project
(COUP) observed \torie\ (COUP 732) for a total exposure of about 10
days over a time period of 3 weeks and found a low level of
variability including one moderate X-ray flare
\citep{Stelzer:al:2005}.  Its luminosity during COUP was determined to
be $\log L_x [\ergsec] = 32.4$; early \chandra\ High Energy
Transmission Grating (HETG) spectra indicated plasma temperatures of
up to 50 MK \citep{Schulz:2003}.

The HETG Orion Legacy Project has now accumulated almost 4 days of
total exposure of \torie\ allowing for an in depth study of its X-ray
spectral properties.  The following analysis of \torie's high
resolution X-ray spectrum is aimed to characterize its coronal nature.
\Vtworev 
The existence of coronal X-rays confirms predictions that very young
intermediate mass stars of less than $4\msun$ are not fully radiative
\citep{Palla93} and may possess some form of magnetic dynamo. We also
compare these properties with the ones observed in \toriaa, various T
\Vtworev 
Tauri stars including the relatively massive T Tauri star, SU~Aur
($2\msun$), active coronal sources, and post-main sequence evolved
G-type giants.  The optical and binary system parameters of \tooe\ can
be found in \citet{Herbig:Griffin:2006}.

\section{Observations and Analysis}
\label{sec:obs}

\subsection{Observations, Data Processing}
\label{subsec:obsproc}

As part of the \hetgs\ Orion Legacy Project, we have observed \tooe\
on 11 separate occasions from 1999 through 2007, mostly within our
\hetg Guaranteed Time program, with individual exposure times ranging
from about 10 to 50 ks.  The \hetgs\ \citep{HETG:2005} is an objective
transmission grating spectrometer with two channels optimized for high
and medium energies (\heg\ and \meg, respectively).  The \heg\ and
\meg\ spectra of each point source in the field form a shallow
``$\times$'' centered on the zeroth order image.  Since the Orion
Trapezium field is crowded, we had to take special care to avoid
source confusion when possible, and to assess contamination and reject
spectra when not.  The range in spacecraft roll angles, the redundancy
provided by multiple gratings and orders, the narrow
point-spread-function, and the efficiency of order-sorting with the
CCD energy resolution all help to provide a reliable spectrum.

We processed the data with \ciao~3.4 \citep{CIAO:2006} taking care to
fine-tune the zero order detection to accurately center on \tooe.
Response files were made with the most recent calibration database
available at the time (version 3.4).  Further analysis was done using
ISIS \citep{Houck:00}, an Interactive Spectral Interpretation System
for high resolution X-ray spectroscopy, developed especially for
scriptable, extensible analysis of \chan\ high resolution spectra.

We give an observing log in Table~\ref{tbl:log}, along with ancillary
information and some derived properties of each observation.
\begin{deluxetable}{rcccrccccc}
  \tabletypesize{\small}
  \tablecolumns{10}
  \tablewidth{0pc}
  \tablecaption{Observation Log}
  \tablehead{
    \colhead{$N$\tablenotemark{a}}&
    \colhead{Date}&
    \colhead{Time}&
    \colhead{$t_{exp}$}&
    \colhead{$\phi$\tablenotemark{b}}&
    \colhead{$C$\tablenotemark{c}}&
    \colhead{Rate$\times 10^3$}&
    \colhead{Flux\tablenotemark{d}$\times10^3$}&
    \colhead{Flux\tablenotemark{d}$\times10^{12}$}&
    \colhead{$v$\tablenotemark{e}}\\
    \colhead{}&
    \colhead{}&
    \colhead{}&
    \colhead{[ks]}&
    \colhead{}&
    \colhead{}&
    \colhead{$\mathrm{[ counts\,s\mone]}$}&
    \colhead{$\mathrm{[ phot\,cm\mtwo\,s\mone]}$}&
    \colhead{$\mathrm{[ ergs\,cm\mtwo\,s\mone]}$}&
    \colhead{$\mathrm{[ km\,s\mone]}$}
  }
  \startdata 
  3   &   1999-10-31&   05:47:21&   49&    0.55&   --   & 105 &1.105 (0.033)&3.30 (0.10)& -11  (22)\\
  4   &   1999-11-24&   05:37:54&   31&    0.93& \sc{h} & 83  &1.250 (0.025)&3.55 (0.07)&  -6  (52)\\
  2567&   2001-12-28&   12:25:56&   46&    0.30&   --   & 69  &0.873 (0.033)&2.27 (0.09)&  65  (55)\\
  2568&   2002-02-19&   20:29:42&   46&    0.70& \sc{hm}& --  &     --      &   --      &      --  \\
  7407&   2006-12-03&   19:07:48&   25&    0.33&   --   & 88  &1.085 (0.052)&3.17 (0.15)&  68  (49)\\
  7410&   2006-12-06&   12:11:37&   13&    0.60&   --   & 77  &1.030 (0.072)&2.87 (0.20)& -4   (72)\\
  7408&   2006-12-19&   14:17:30&   25&    0.93&   --   & 78  &0.994 (0.049)&2.88 (0.14)& -35  (60)\\
  7409&   2006-12-23&   00:47:40&   27&    0.28&   --   & 90  &1.140 (0.050)&3.43 (0.15)&  96  (40)\\
  8897&   2007-11-15&   10:03:16&   24&    0.35&   --   & 84  &1.065 (0.052)&3.16 (0.15)&  72  (41)\\
  8896&   2007-11-30&   21:58:33&   23&    0.93& \sc{m} & 25  &1.030 (0.043)&3.00 (0.13)& -7   (82)\\
  8895&   2007-12-07&   03:14:07&   25&    0.55&   --   & 86  &1.045 (0.049)&3.10 (0.15)&  86  (37)\\
  \enddata
  \tablenotetext{a}{$N$ is the \chan observation identifier number.}
  %
  \tablenotetext{b}{Orbital phase ($\phi$) was computed from the ephemeris of
    \citet{Herbig:Griffin:2006}.}  
  \tablenotetext{c}{The ``$C$'' column indicates spectra with
    severe confusion, ``H'' and ``M'' respectively indicate whether
    the \heg\  or \meg\ spectrum could not be used.}
  %
  \tablenotetext{d}{The model-independent average flux in the
    $2$--$17$\AA\ range, computed from \heg and \meg counts,
    flux-corrected using the responses.}
  \tablenotetext{e}{The line-of-sight velocity, heliocentric
    correction applied. Values in parentheses are the 90\%
    errorbar ($\sim1.6\sigma$). See Section~\ref{sec:dyn} for explanation.}
  \label{tbl:log}
\end{deluxetable}


\subsection{Source Confusion}
\label{subsec:conf}

To assess confusion in detail, we used two techniques.  For field
source zeroth-order coincidence with the diffracted spectra of \tooe,
we used the COUP \citep{Getman:Flaccomio:al:2005} source list, and for
each observation we transformed the list's celestial coordinates to
the diffraction coordinate system for \tooe. We did not find any
source on the spectral regions with significant counts.  
\note{sentence deleted}
%

The second technique assessed the contamination from sources near the
zeroth order, such that their diffracted spectra would overlap with
the \heg\ or \meg\ spectra of \tooe, and so close to the zeroth order
that the order-sorting by CCD energy would distinguish orders.  For
this, it was crucial to inspect the events' distribution as selected
from the \tooe\ default binning region (cross-dispersion region half
width of $6.6\times10\mfour\,\mathrm{deg}$) in diffraction distance
versus energy (the CCD blurred energy) coordinates in which zeroth
orders appear as vertical distributions and diffracted photons as
hyperbolas.  Here we found significant contamination for a few
observations and had to reject all or some orders.

\begin{figure}[h]
  \includegraphics{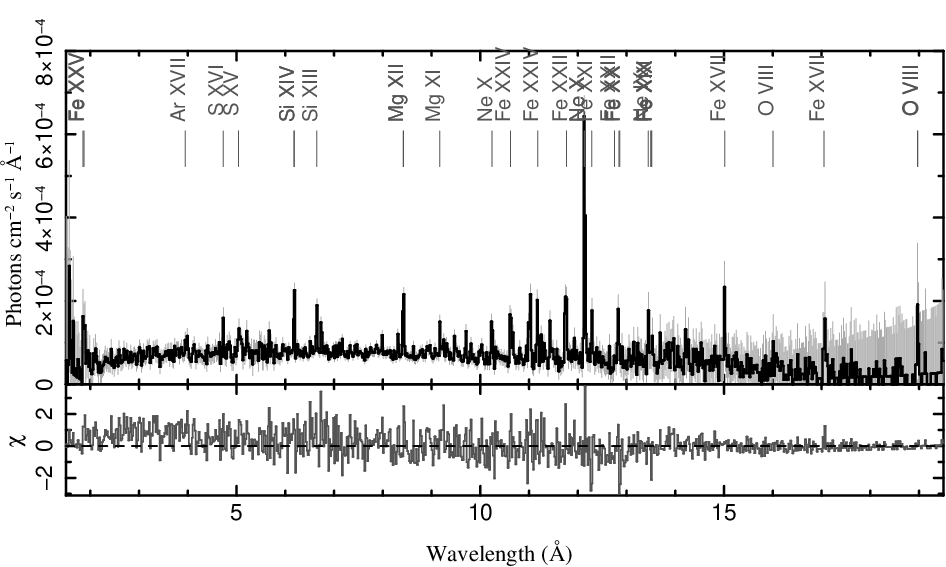} 
  \caption{Here is an overview of the cumulative spectrum (about 260
    ks) for combined \meg and \heg flux, for a bin size of 0.02
    \AA. Some significant lines are marked.  The statistical counting
    uncertainty is shown in gray. The lower panel shows the residuals
    against the emission measure model.  The integrated flux
    ($1$--$40$ \AA) at Earth is
    $3.3\times10^{-12}\,\mathrm{ergs\,cm\mtwo s\mone}$
    ($1.3\times10^{-3}\,\mathrm{phot\,cm\mtwo s\mone}$).
}
  \label{fig:overviewspec}
\end{figure}

The useful exposure from which we can extract spectra, light curves,
and line fluxes totals to 260 ks.  Rejected orders or observations are
flagged in Table~\ref{tbl:log}.  A cumulative counts spectrum is shown
in Figure~\ref{fig:overviewspec}.
%

\subsection{Light Curves, Variability}\label{subsec:var}

\begin{figure}[h]
  \includegraphics[angle=0,scale=1.0]{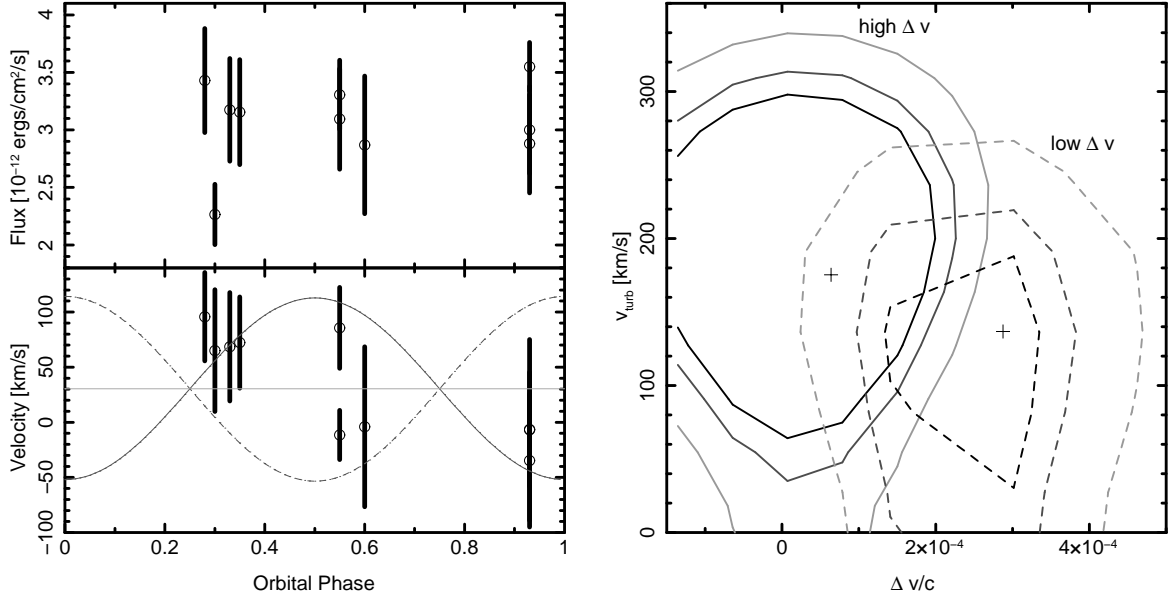} 
  \caption{Left panel, top: the circles show the mean fluxes of the
    \heg and \meg spectra over the wavelength range 2--17 \ang for
    each unconfused observation (see Table~\ref{tbl:log}). Error-bars
    show $3\sigma$ uncertainties.  Left panel, bottom: circles show
    the measured radial velocities and 90\% uncertainties.  Stellar
    component radial velocities (solid and dashed lines in the lower
    panel) and orbital phases were computed from the ephemeris and
    orbital solution of \citet{Herbig:Griffin:2006}. Right panel:
    $\chi^2$ confidence countours ($1\sigma$, $2\sigma$, and
    $3\sigma$, inner to outer) for Doppler shift ($\Delta v/c$)
    against the turbulent broadening parameter.  These were computed
    for two groups of spectra, those nearest the minimum orbital
    radial velocity separation (``low $\Delta v$''; dashed), and those
    nearest the maximum (``high $\Delta v$''; solid).  }
  \label{fig:fluxlc}
\end{figure}

There was little variability of any significance within any
observation. The \meg\ rate was about 0.05 counts/s.  Variations
within each observation were consistent with statistical uncertainties
--- no abrupt increases or slow decays characteristic of coronal
flares occurred.  Observation to observation, there was also no
variability, except for one which had a significantly lower count rate
than the others.  We show the mean flux rate per observation for \meg\
and \heg\ spectra from $2$--$17$ \AA\ in the upper left graph in
Figure~\ref{fig:fluxlc}, phased using the ephemeris of
\citet{Herbig:Griffin:2006}; values are also listed in
Table~\ref{tbl:log}.
\Vtworev 
\Vtworev 

\subsection{Spectral Analysis}\label{subsec:spec}

Spectral analysis is an iterative process.  For detailed spectral
diagnostics and models, accurate line fluxes, centroids, and widths
are fundamental to determinations of temperature structure,
abundances, and dynamics.  However even at \hetgs\ resolution, we are
dependent upon plasma models for accurate estimation of the continuum
and for assessment of blending.  We base our spectral models on the
Astrophysical Plasma Emission Database \citep[\aped;][]{Smith:01},
ionization balance of \citet{Mazzotta:98}, and abundances of
\citet{Anders:89}.  We begin by fitting a one-temperature component
model to the short wavelength continuum, then add one or two more
components to get a reasonable match to the continuum throughout the
spectrum, ignoring strong lines in the process.  Next we fit about 100
lines parametrically with unresolved Gaussians folded through the
instrumental response, using the line-free plasma model for the local
continuum.  To improve the statistic per bin, we group the spectra by
\Vtworev
2--4 bins and we combined the \meg\ first orders and combined the
\heg\ first orders, both over all observations, then fit the \meg\ and \heg\
jointly.  (The default binning oversamples the \heg and \meg
resolutions by a factor of two, or 0.0025 \AA\ and 0.005 \AA,
respectively.)  Combination of spectra is done dynamically --- each
effective area and redistribution matrix are distinct, and summed
counts are compared with the summed folded models to compute the
statistic.  We adopted an interstellar absorption column of $N_H =
2\times10^{21}\,\mathrm{cm\mtwo}$ as determined by
\citet{Schulz:Canizares:al:2001}.  For some weak lines, we froze the
wavelength at the theoretical value.  This allows us to obtain a limit
on the flux which can provide important constraints on emission
measure and abundance reconstruction.

After we have mean line fluxes and centroids for a variety of elements
and ions (see Table~\ref{tbl:fluxes}), we can reconstruct the emission
measure distribution assuming that the plasma has uniform abundances
and is in collisional ionization equilibrium.  We use a uniform
logarithmic temperature grid and minimize the line flux residuals by
adjusting the weights in each temperature bin as well as the elemental
abundances.  Since this is an ill-conditioned problem, we impose a
smoothness on the emission measure by using its sum-squared second
derivative in a penalty function.  

\note{``Emission measure...'' updated}
Emission measure reconstruction is also an iterative process.  First
we ignore lines with large wavelength residuals relative to their
preliminary identification based on expectations of a baseline plasma
model, since they are likely misidentified or blended.  Then we
reconstruct a trial emission measure distribution. Lines with large
flux residuals are rejected, since they may be symptomatic of
unresolved blends or inaccurate emissivities due to uncertainties in
the underlying atomic data. We repeat the fit with the accepted
lines.  We use the emission measure and abundance model to generate a
synthetic spectrum and compare to the observed spectrum.  Here we can
adjust the line-to-continuum ratio by adjusting the normalizations of
the emission measure and relative abundances.  If the continuum model
was improved, we start over by fitting the lines with the improved
continuum.  Finally, we perform a Monte-Carlo series of fits in which
we let the measured line flux vary randomly according to its measured
uncertainty.  This provides an estimate of the uncertainty on the
emission measure and the abundances.

\clearpage
%
%
%
%
%
%
%
%

\begin{deluxetable}{rccccc}
  \tablecolumns{6}
  \tablewidth{6in}
  \tablecaption{Line Flux Measurements}
  \tablehead{
    \colhead{Ion}&
    \colhead{$\log T_{\mathrm{max}}$}&
    \colhead{$\lambda_{\mathrm{pred}}$}&
    \colhead{$\lambda_{\mathrm{meas}}$\tablenotemark{a}}&
    \colhead{$f_{\mathrm{meas}}$\tablenotemark{a}}&
    \colhead{$f_{\mathrm{pred}}$}\\
    %
    \colhead{}&
    \colhead{$[\log\mathrm{K}]$}&
    \colhead{[\AA]}&
    \colhead{[\AA]}&
    \multicolumn{2}{c}{$[10^{-6}\,\mathrm{phot\,cm\mtwo\,s\mone}]$}
  }
  \startdata
  \eli{Fe}{25}&  7.8&   1.8607&   1.8684 (0.0073)&    3.293 (1.846)&    1.433\\ 
  \eli{Ca}{19}&  7.5&   3.1772&   3.1769 (0.0023)&    1.135 (0.421)&    0.518\\ 
  \eli{Ca}{19}&  7.4&   3.1909&   3.1909 (0.0000)&    0.033 (0.223)&    0.172\\ 
  \eli{Ca}{19}&  7.5&   3.2110&   3.2110 (0.0000)&    0.373 (0.355)&    0.156\\ 
  \eli{Ar}{18}&  7.7&   3.7338&   3.7355 (0.0081)&    0.328 (0.351)&    0.236\\ 
  \eli{Ar}{17}&  7.4&   3.9491&   3.9582 (0.0059)&    0.961 (0.558)&    0.397\\ 
   \eli{S}{15}&  7.3&   4.0883&   4.0883 (0.0000)&    0.093 (0.226)&    0.049\\  
   \eli{S}{16}&  7.6&   4.7301&   4.7272 (0.0026)&    1.615 (0.530)&    1.030\\  
  \eli{Si}{14}&  7.4&   4.9468&   4.9468 (0.0000)&    0.193 (0.311)&    0.124\\ 
   \eli{S}{15}&  7.2&   5.0387&   5.0405 (0.0035)&    1.253 (0.556)&    1.136\\  
   \eli{S}{15}&  7.2&   5.0648&   5.0648 (0.0000)&    0.541 (0.499)&    0.246\\  
   \eli{S}{15}&  7.2&   5.1015&   5.1038 (0.0051)&    0.991 (0.505)&    0.373\\  
  \eli{Si}{14}&  7.4&   5.2174&   5.2172 (0.0065)&    1.371 (0.604)&    0.580\\ 
  \eli{Si}{13}&  7.1&   5.4045&   5.4045 (0.0000)&    0.031 (0.255)&    0.131\\ 
  \eli{Si}{13}&  7.1&   5.6805&   5.6698 (0.0030)&    1.507 (0.511)&    0.391\\ 
  \eli{Si}{13}&  6.9&   5.8160&   5.8042 (0.0035)&    0.815 (0.410)&    0.027\\ 
  \eli{Si}{14}&  7.4&   6.1831&   6.1825 (0.0008)&    4.908 (0.466)&    4.143\\ 
  \eli{Si}{13}&  7.0&   6.6480&   6.6469 (0.0011)&    3.341 (0.426)&    2.937\\ 
  \eli{Mg}{12}&  7.2&   7.1063&   7.1028 (0.0018)&    1.095 (0.290)&    0.735\\ 
  \eli{Mg}{11}&  6.9&   7.3101&   7.3101 (0.0000)&    0.173 (0.211)&    0.044\\ 
  \eli{Fe}{22}&  7.1&   7.6812&   7.6812 (0.0000)&    0.026 (0.136)&    0.073\\ 
  \eli{Al}{12}&  7.0&   7.7573&   7.7668 (0.0054)&    0.521 (0.281)&    0.458\\ 
  \eli{Mg}{11}&  6.9&   7.8503&   7.8503 (0.0000)&    0.472 (0.277)&    0.257\\ 
  \eli{Al}{12}&  6.9&   7.8721&   7.8721 (0.0000)&    0.608 (0.300)&    0.319\\ 
  \eli{Fe}{23}&  7.2&   7.9009&   7.9009 (0.0000)&    0.079 (0.181)&    0.149\\ 
  \eli{Fe}{24}&  7.4&   7.9857&   7.9794 (0.0138)&    0.515 (0.484)&    0.523\\ 
  \eli{Fe}{24}&  7.4&   7.9960&   7.9894 (0.0121)&    1.055 (0.733)&    0.266\\ 
  \eli{Fe}{23}&  7.2&   8.3038&   8.3055 (0.0025)&    1.078 (0.346)&    0.552\\ 
  \eli{Fe}{24}&  7.4&   8.3161&   8.3217 (0.0022)&    0.901 (0.335)&    0.581\\ 
  \eli{Fe}{24}&  7.4&   8.3761&   8.3810 (0.0037)&    0.321 (0.272)&    0.225\\ 
  \eli{Mg}{12}&  7.2&   8.4219&   8.4215 (0.0008)&    5.875 (0.531)&    5.198\\ 
  \eli{Fe}{21}&  7.1&   8.5740&   8.5643 (0.0056)&    0.377 (0.306)&    0.295\\ 
  \eli{Fe}{23}&  7.2&   8.8149&   8.8158 (0.0062)&    0.357 (0.302)&    0.566\\ 
  \eli{Fe}{22}&  7.1&   8.9748&   8.9741 (0.0047)&    0.811 (0.360)&    0.582\\ 
  \eli{Mg}{11}&  6.8&   9.1687&   9.1710 (0.0015)&    2.590 (0.475)&    1.866\\ 
  \eli{Fe}{21}&  7.1&   9.1944&   9.1944 (0.0000)&    0.249 (0.307)&    0.220\\ 
  \eli{Fe}{22}&  7.1&   9.3933&   9.3865 (0.0150)&    0.277 (0.355)&    0.126\\ 
  \eli{Ne}{10}&  7.0&   9.7083&   9.7082 (0.0048)&    2.002 (0.899)&    1.210\\ 
  \eli{Ni}{19}&  6.8&  10.1100&  10.1169 (0.0150)&    0.018 (0.220)&    0.044\\ 
  \eli{Fe}{20}&  7.0&  10.1203&  10.1332 (0.0032)&    0.654 (0.432)&    0.194\\ 
  \eli{Ne}{10}&  7.0&  10.2390&  10.2390 (0.0015)&    4.084 (0.634)&    3.736\\ 
  \eli{Fe}{24}&  7.4&  10.6190&  10.6166 (0.0041)&    3.659 (1.812)&    3.792\\ 
  \eli{Fe}{19}&  6.9&  10.6491&  10.6435 (0.0086)&    0.759 (0.764)&    0.237\\ 
  \eli{Fe}{24}&  7.4&  10.6630&  10.6650 (0.0027)&    2.314 (0.652)&    1.990\\ 
  \eli{Fe}{23}&  7.2&  11.0190&  11.0188 (0.0043)&    2.517 (1.651)&    1.944\\ 
  \eli{Fe}{24}&  7.4&  11.0290&  11.0312 (0.0039)&    3.934 (1.563)&    2.460\\ 
  \eli{Fe}{24}&  7.4&  11.1760&  11.1764 (0.0015)&    6.469 (0.946)&    4.445\\ 
  \eli{Fe}{18}&  6.8&  11.3260&  11.3132 (0.0041)&    1.337 (0.574)&    0.469\\ 
   \eli{Ne}{9}&  6.6&  11.5440&  11.5436 (0.0060)&    0.578 (0.519)&    0.560\\  
  \eli{Fe}{23}&  7.2&  11.7360&  11.7400 (0.0012)&    7.656 (1.068)&    6.279\\ 
  \eli{Fe}{22}&  7.1&  11.7700&  11.7719 (0.0014)&    6.244 (0.996)&    5.384\\ 
  \eli{Ne}{10}&  6.9&  12.1348&  12.1352 (0.0010)&   30.490 (2.300)&   25.433\\ 
  \eli{Fe}{23}&  7.2&  12.1610&  12.1610 (0.0000)&    4.296 (1.306)&    3.492\\ 
  \eli{Fe}{17}&  6.7&  12.2660&  12.2713 (0.0053)&    2.024 (0.937)&    0.900\\ 
  \eli{Fe}{20}&  7.0&  13.3850&  13.3844 (0.0075)&    1.416 (1.071)&    0.962\\ 
  \eli{Fe}{19}&  6.9&  13.4230&  13.4230 (0.0000)&    1.162 (1.131)&    0.389\\ 
   \eli{Ne}{9}&  6.6&  13.4473&  13.4493 (0.0050)&    5.083 (2.097)&    3.874\\  
  \eli{Fe}{19}&  6.9&  13.4620&  13.4620 (0.0000)&    0.893 (1.591)&    0.883\\ 
  \eli{Fe}{19}&  6.9&  13.4970&  13.4970 (0.0000)&    2.554 (1.311)&    1.552\\ 
  \eli{Fe}{19}&  6.9&  13.5180&  13.5324 (0.0048)&    4.074 (1.938)&    3.422\\ 
  \eli{Fe}{19}&  6.9&  13.6450&  13.6599 (0.0150)&    0.609 (0.889)&    0.543\\ 
   \eli{Ne}{9}&  6.6&  13.6990&  13.6927 (0.0043)&    3.873 (1.503)&    1.714\\  
  \eli{Fe}{19}&  6.9&  13.7950&  13.7986 (0.0109)&    1.971 (1.789)&    1.361\\ 
  \eli{Fe}{17}&  6.7&  13.8250&  13.8351 (0.0052)&    3.759 (1.585)&    0.718\\ 
  \eli{Fe}{18}&  6.8&  14.2080&  14.2132 (0.0037)&    6.378 (1.987)&    5.219\\ 
  \eli{Fe}{18}&  6.8&  14.2560&  14.2560 (0.0000)&    0.442 (1.192)&    0.996\\ 
  \eli{Fe}{20}&  7.0&  14.2670&  14.2670 (0.0000)&    3.975 (2.142)&    1.300\\ 
  \eli{Fe}{18}&  6.8&  14.3430&  14.3430 (0.0000)&    0.323 (0.727)&    0.618\\ 
  \eli{Fe}{18}&  6.8&  14.3730&  14.3730 (0.0000)&    1.205 (1.127)&    1.325\\ 
  \eli{Fe}{18}&  6.8&  14.5340&  14.5355 (0.0050)&    3.860 (1.546)&    1.003\\ 
  \eli{Fe}{17}&  6.7&  15.0140&  15.0109 (0.0022)&   10.500 (2.349)&    8.749\\ 
  \eli{Fe}{19}&  6.9&  15.0790&  15.0825 (0.0068)&    2.224 (1.393)&    1.098\\ 
    \eli{O}{8}&  6.7&  15.1762&  15.1680 (0.0075)&    1.598 (1.276)&    0.664\\   
  \eli{Fe}{19}&  6.9&  15.1980&  15.2082 (0.0058)&    2.238 (1.398)&    0.928\\ 
  \eli{Fe}{17}&  6.7&  15.2610&  15.2600 (0.0061)&    2.795 (1.488)&    2.462\\ 
  \eli{Fe}{18}&  6.8&  15.8700&  15.8572 (0.0039)&    2.800 (1.661)&    0.432\\ 
  \eli{Fe}{18}&  6.8&  16.0710&  16.0814 (0.0058)&    2.960 (1.796)&    1.813\\ 
  \eli{Fe}{17}&  6.7&  16.7800&  16.7650 (0.0029)&    3.114 (2.137)&    3.792\\ 
  \eli{Fe}{17}&  6.7&  17.0510&  17.0547 (0.0041)&   10.810 (3.391)&    4.557\\ 
  \eli{Fe}{17}&  6.7&  17.0960&  17.1011 (0.0048)&    6.809 (2.930)&    4.086\\ 
  \eli{Fe}{18}&  6.8&  17.6230&  17.6080 (0.0025)&    4.326 (2.586)&    1.323\\ 
    \eli{O}{7}&  6.4&  17.7680&  17.7680 (0.0000)&    0.300 (0.315)&    0.041\\   
    \eli{O}{7}&  6.4&  18.6270&  18.6270 (0.0000)&    0.900 (0.926)&    0.123\\   
    \eli{O}{8}&  6.7&  18.9698&  18.9744 (0.0043)&   17.680 (5.846)&   13.292\\   
  \enddata
  \label{tbl:fluxes}
  \tablecomments{These are the lines which were used in the emission
    measure and abundance reconstruction.  The $f_\mathrm{pred}$, is
    what the emission measure and abundance model predicts.}
  \tablenotetext{a}{Values in parentheses are $1\sigma$ uncertainties.
    If the uncertainty for the wavelength is $0.0$, then the line
    position was frozen in the fit.}
\end{deluxetable}

\clearpage

\begin{figure}[htb]
  \includegraphics[scale=0.67]{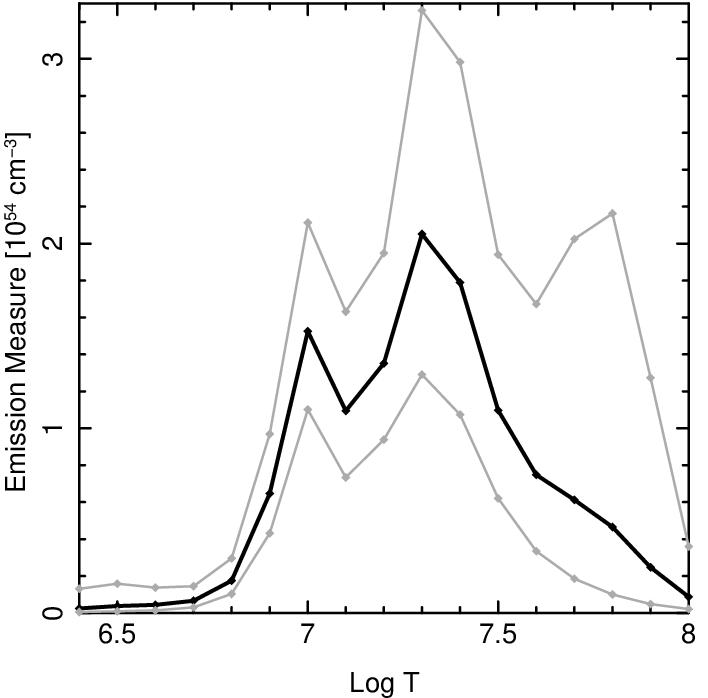} 
  \caption{Reconstructed emission measure, for a distance of 450 pc
    and $N_H = 2\times10^{21}\mathrm{cm\mtwo}$.  The upper and lower
    gray boundaries are $1\sigma$ (logarithmic) statistical
    uncertainties from Monte-Carlo iterations in which the measured
    fluxes were perturbed randomly according to their measurement
    uncertainties.  The integrated emission measure is
     $1.2\times10^{55}\,\mathrm{cm\mthree}$. }
  \label{fig:dem}
\end{figure}

\begin{figure}[htb]
  \includegraphics[scale=0.67]{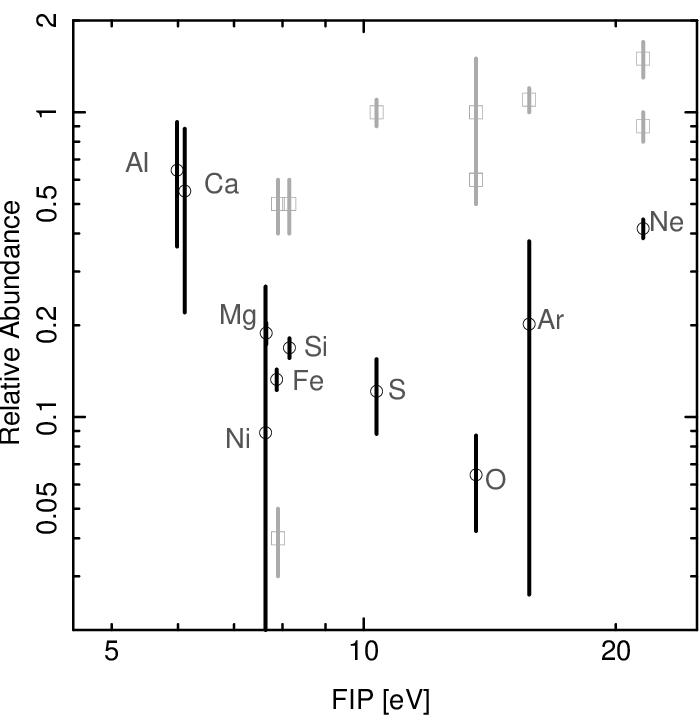}  
  \caption{\torie abundances (circles) relative to Solar photospheric
    values \citep{Anders:89} as derived from emission measure
    reconstruction. Error-bars give the statistical $1\sigma$
    uncertainty.  Light colored squares give Orion stellar
    photospheric
    \citep{Cunha:Hubeny:al:2006,Cunha:Lambert:1994,Cunha:Smith:al:1998,Cunha:Smith:2005}
    or nebular \citep{Esteban:al:2004} abundance ratios for Fe, Si, S,
    O, Ar, and Ne. }
  \label{fig:abund}
\end{figure}

We have applied this technique to several other spectra \citep[e.g.,
see][ and references therein]{Huenemoerder:Kastner:al:2007}.  Given
the form of the problem there is no unique solution.  However, results
can be useful for comparison of emission measures derived with similar
methods.  Figure~\ref{fig:dem} shows our reconstructed emission
measure distribution whose values are also given in
Table~\ref{tbl:emeas}.  The corresponding abundances are shown in
Figure~\ref{fig:abund} and are listed in Table~\ref{tbl:abund}.

\clearpage
 \begin{deluxetable}{cccc}
  \tablecolumns{4}
  \tablewidth{0in}
  \tablecaption{Emission Measure Model}
  \tablehead{
    \colhead{$\log T$}&
    \colhead{$EM$}&
    \colhead{$EM_{low}$}&
    \colhead{$EM_{high}$}\\
    \colhead{$[\log K]$}&
    \colhead{}&
    \colhead{$\mathrm{[10^{54} cm\mthree]}$}&
    \colhead{}
  }
  \startdata
 6.4&   2.45e-02&   4.60e-03&   1.31e-01\\
 6.5&   3.76e-02&   8.90e-03&   1.59e-01\\
 6.6&   4.40e-02&   1.41e-02&   1.37e-01\\
 6.7&   6.59e-02&   3.00e-02&   1.45e-01\\
 6.8&   1.74e-01&   1.03e-01&   2.96e-01\\
 6.9&   6.46e-01&   4.31e-01&   9.69e-01\\
 7.0&   1.53e+00&   1.10e+00&   2.11e+00\\
 7.1&   1.09e+00&   7.33e-01&   1.63e+00\\
 7.2&   1.35e+00&   9.38e-01&   1.95e+00\\
 7.3&   2.05e+00&   1.29e+00&   3.26e+00\\
 7.4&   1.79e+00&   1.07e+00&   2.98e+00\\
 7.5&   1.10e+00&   6.21e-01&   1.94e+00\\
 7.6&   7.48e-01&   3.35e-01&   1.67e+00\\
 7.7&   6.13e-01&   1.85e-01&   2.02e+00\\
 7.8&   4.65e-01&   1.00e-01&   2.16e+00\\
 7.9&   2.47e-01&   4.79e-02&   1.27e+00\\
 8.0&   8.67e-02&   2.09e-02&   3.60e-01\\
 \enddata
 \label{tbl:emeas}
 \tablecomments{The reconstructed emission measure over the
   temperature range of sensitive features.  The ``low'' and ``high''
   values are the logarithmic $1\sigma$ boundaries from Monte-Carlo
   iterations.} 
\end{deluxetable}
 \begin{deluxetable}{ccc}
  \tablecolumns{3}
  \tablewidth{3.5in}
  \tablecaption{Relative Elemental Abundances}
  \tablehead{
    \colhead{Element}&
    \colhead{Abund\tablenotemark{a}}&
    \colhead{FIP\tablenotemark{b}}\\
    \colhead{}&
    \colhead{}&
    \colhead{[eV]}
  }

  \startdata
  O &    0.06  (0.03)&    13.618\\
  Ne&    0.42  (0.03)&    21.564\\
  Mg&    0.19  (0.02)&     7.646\\
  Al&    0.64  (0.28)&     5.986\\
  Si&    0.17  (0.01)&     8.151\\
  S &    0.12  (0.03)&    10.360\\
  Ar&    0.20  (0.17)&    15.759\\
  Ca&    0.55  (0.33)&     6.113\\
  Fe&    0.13  (0.01)&     7.870\\
  Ni&    0.14  (0.14)&     7.635\\
  \enddata
  \label{tbl:abund}
  \tablecomments{Uncertainties in ``()'' are $1\sigma$ values.}
  \tablenotetext{a}{Abundances are relative to Solar using the values of \citet{Anders:89}}
  \tablenotetext{b}{FIP is the first ionization potential.}
\end{deluxetable}

\subsubsection{Dynamics}\label{sec:dyn}

\Vtworev 

We searched for dynamical effects by using the mean plasma model
spectrum as a template for fitting narrow regions by adjusting the
Doppler shift, turbulent broadening velocity, and local
normalizations.  The 10--12.5 \AA\ region has a number of lines which
make it useful for this purpose,
\Vtworev 
and we used that entire wavelength interval in the fits.
We fit each observation independently for line-of-sight velocity.  The
results are listed in Table~\ref{tbl:log} and shown in
Figure~\ref{fig:fluxlc}.  The radial velocities are consistent with
the orbital velocities of the stellar components --- that is, they are
generally less than or equal to the orbital radial velocities, but are
limited by the spectrometer's sensitivity.  According to the orbital
solution of \citet{Herbig:Griffin:2006}, the stellar components have a
projected radial velocity amplitude of about $80\,\mathrm{km\,s\mone}$
and a systemic velocity of $30\, \mathrm{km\,s\mone}$ (note that for
the adopted masses of $3.5 \msun$, the inclination is $i=61^\circ$).
While there is a slight systematic offset towards the red-shifted
component at some phases (e.g., 0.28, 0.30, 0.33, 0.35), the trend
does not persist since at other phases (e.g., 0.55, 0.60),
measurements span the range between the components' orbital
velocities.

To obtain better sensitivity and to examine line widths, we also fit
spectra combined into two groups, one nearest to zero velocity
separation (near phase 0.25; 4 observations) and the other nearest to
maximum velocity separation (near phase 0.5 or 1.0; 6 observations).
We computed contour maps in line-of-sight and turbulent broadening
velocities.  The point of this is not that we necessarily expect
turbulent broadening, but that there could be broadening due to binary
orbital effects, and fitting turbulent broadening is simply a useful
parameterization of this.  For instance, for equally X-ray bright
stars, the X-ray line's measured radial velocity could always be zero
(or the systemic value), but the lines could broaden and narrow,
modulated by the orbital radial velocity.  If only one star were the
X-ray source, the lines could shift but maintain constant width.
Since we are photon-limited, we need to group spectra in order to
obtain significant counting statistics.

The resulting confidence contours are shown in
Figure~\ref{fig:fluxlc}.  We see the radial velocity offset towards a
small positive velocity as we should in the low-$\Delta v$ group at a
level of about 30--120 $\mathrm{km\,s\mone}$ (or $\Delta v/c\sim 1$--$4
\times 10^{-4};$ 90\% confidence limits), while the high-$\Delta v$
group range is $-60$ to $60\, \mathrm{km\,s\mone}$.  Broadening is
marginally significant with 90\% confidence contours of $60$ -- $300
\,\mathrm{km\,s\mone}$ for the high-$\Delta v$ group and $0$ --
$200\,\mathrm{km\,s\mone}$ for the low-$\Delta v$ group (with a
$1\sigma$ lower limit of $20\,\mathrm{km\,s\mone}$).

\Vtworev 

\subsubsection{He-like Triplets}

The \hetgs\ bandpass includes the He-like triplet lines of \mgxi,
\neix, and \ovii, which are useful diagnostics of density in the
coronal regime \citep{Gabriel:69,Gabriel:73}.  Due to the absorption
towards Orion and the low sensitivity of the \hetgs\ at 22 \AA, we
have no useful data on \ovii, but we do have spectra of \mgxi\ and
\neix.  By fitting the line ratios of the resonance ($r$),
intercombination ($i$), and forbidden ($f$) lines we are able to put
upper limits on the coronal electron density.  In our fits, we
constrained the relative positions of the triplet components and
included blends as estimated from the emission measure model spectrum.
\Vtworev 
The \eli{Ne}{10} Lyman series converges near the \mgxi\ $i$-line
(9.230 \AA). We included relevant lines of this series by providing an
initial guess for their strengths by scaling fluxes according to their
relative $f$-values from the isolated and well detected Ly-$\gamma$
and $\delta$ lines, since APED does not contain lines with upper
levels $n>5$.  The locations of the weaker and blended features are
9.215 \AA\ (\eli{Ne}{10} Ly-$\alpha$ series $n=9$ to $n=1$
transition), 9.246 \AA\ (\eli{Ne}{10} 8-to-1 transition), and
9.194 \AA\ (\eli{Fe}{21}).  The weakest (\eli{Fe}{21}) line's
position was frozen, while the others positions and fluxes were left
free and gave reasonable fitted wavelengths.  While this is not a
complete and accurate plasma model for the region, it is a useful
parameterization to obtain ratios for the interesting features.

There are Fe blends in the \neix\ triplet region, but since the Ne:Fe
abundance ratio is high (see Table~\ref{tbl:abund}), these are not
severe.

The continuum in each region was evaluated from the plasma model and
not governed by free parameters.

\begin{figure}[h]
  \includegraphics[angle=0,scale=0.5]{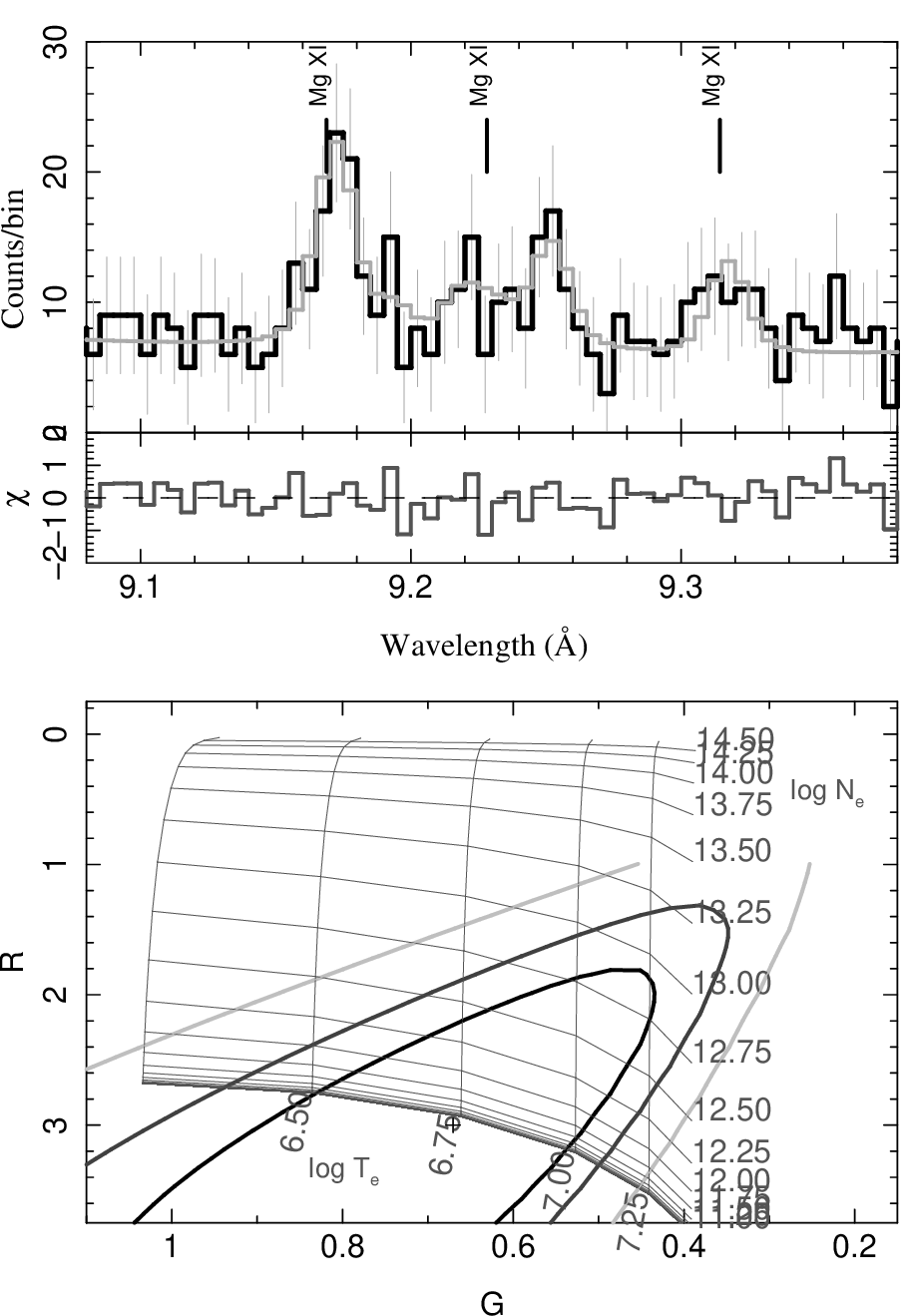}  
  \hspace{2mm}
  \includegraphics[angle=0,scale=0.5]{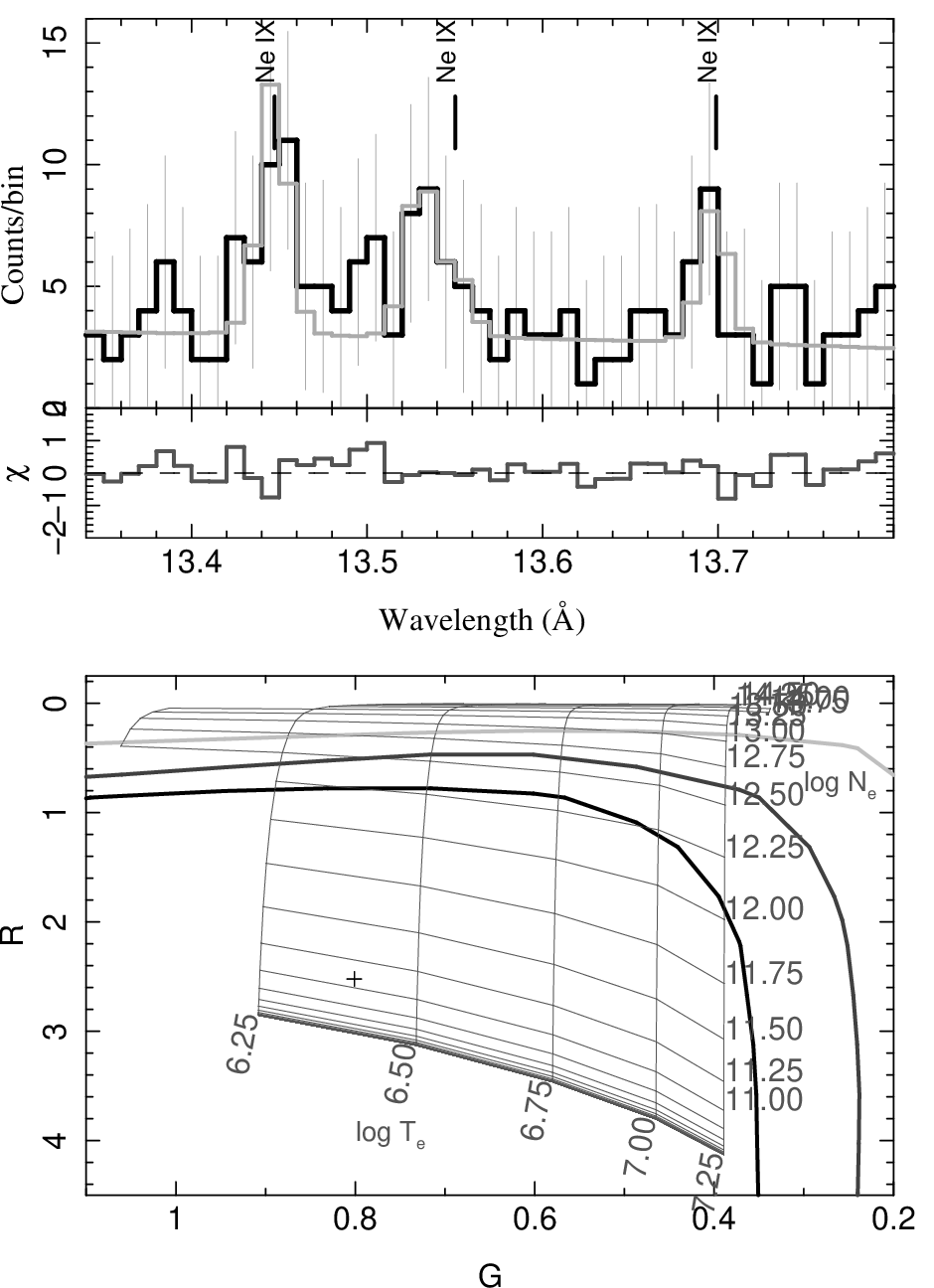}   
  \caption{The upper plots show the \mgxi\ (left) and \neix\ (right)
    triplets counts spectra (black histogram), a multi-Gaussian fit
    (gray line) and residuals (lower panels of top plots).  The
    positions of the triplet components are marked.  The lower plots
    show the fits to the $G$ ($=(f+i)/r$) and $R$ ($=f/i$) ratios, which
    are primarily temperature and density sensitive, respectively, as
    given by the grid.  The contours are the $1\sigma$, $2\sigma$, and
    $3\sigma$ levels from lower/inner to upper/outer contours.}
  \label{fig:triplets}
\end{figure}

\Vtworev 
Results for both triplets are consistent with low density.  
In Figure~\ref{fig:triplets} we show the spectra and confidence
contours of the $G$ and $R$ ratios, defined as $G=(f+i)/r$ and
$R=f/i$.  $G$ is primarily a function of temperature, and $R$ of
density. The density and temperature dependence are from APEC
calculations \citep{Smith:01}.\footnote{Data for the lines are
  available from {\small\tt
    http://cxc.harvard.edu/atomdb/features/denHETG.ps}}

%

\clearpage
\subsubsection{Abundance Ratios}
By forming appropriately weighted flux ratios of line pairs, we can
obtain relatively temperature-insensitive abundance ratios.  This is
achieved by relating a linear combination of emissivities of He and
H-like resonance lines for one element to a similar linear combination
of another element, so as to minimize fluctuations with temperature in
their ratio. If this ratio is constant, then the analogous combination
of line fluxes gives the abundance ratio.  This technique was
explained in detail by \citet{Liefke:Ness:al:2008} and described generally
by \citet{Garcia-Alvarez:al:2005}.  

\note{updated notation}
We define the ratio in the following way:
\begin{equation}
  \label{eq:tinsensratio}
  r = 
  \frac{  F_{1,1} + a_1 F_{1,2}  } 
  {   F_{2,1} + a_2 F_{2,2}  }
  =
  \frac{ A_1 a_0}{ A_2}
  \frac{ \int{ [\epsilon_{1,1}(T) + a_1 \epsilon_{1,2}(T) ] D(T)\, d T}}
  { \int{ [\epsilon_{2,1}(T) + a_2 \epsilon_{2,2}(T) ] D(T)\, d T}}
\end{equation}
in which subscripts ${i,j}$ on $F$ refer to the line $j$ from element
$i$, $F$ is the observed flux, $\epsilon(T)$ is the emissivity, $D(T)$
the emission measure, and $A_i$ the abundance of element $i$.  The
parameters, $a_n$ are to-be-determined. It is clear that if the terms
in square brackets within the integrals are identical in numerator and
denominator for all $T$, then the integrals cancel and we are left
with the abundance ratio (parameters $a_1$ and $a_2$ serve to flatten
the ratio, and $a_0$ normalizes it). We determined the parameters by
minimizing the variation in the ratios using H-like and He-like line
emissivities from the APED database.  Re-writing in terms of the
abundance ratio for these lines, we have
\begin{equation}
  \label{eq:tinsensfinal}
  \frac{A_1}{A_2} = \left(\frac{1}{a_0}\right)
   \frac{  F_{1,H} + a_1 F_{1,He}  } 
          {F_{2,H} + a_2 F_{2,He}  } 
\end{equation}
in which subscripts $H$ and $He$ represent the hydrogen-like
Lyman-$\alpha$ doublet and the helium-like Lyman-$\alpha$ resonance
line, respectively.  We tabulate coefficients for a few useful ratios
in Table~\ref{tbl:tinsens} as derived from APED for units of $F$ in
$[\mathrm{phot\,cm\mtwo s\mone}]$ and abundances relative to Solar.
Minimization of the ratio variances were restricted to temperature
ranges where the emissivities were greater than 1\% of their maximum.
In Table~\ref{tbl:abundratios} we give the abundance ratios derived
from the temperature-insensitive ratios along with values from
emission measure modeling.  The ratios from each method are in very
good agreement.  This means that abundance ratios can be derived
fairly easily, without resorting to emission measure reconstruction.
 \begin{deluxetable}{rcccc}
  \tablecolumns{5}
  \tablewidth{4.0in}
  \tablecaption{Temperature Insensitive Ratio Coefficients}
  \tablehead{
    \colhead{Ratio}&
    \colhead{$a_0$}&
    \colhead{$a_1$}&
    \colhead{$a_2$}&
    \colhead{$\sigma$}
  }
  \startdata
  Ne:Mg&  1.440& 0.050& 2.675& 0.273\\
  Mg:Si&  0.536& 0.085& 1.920& 0.064\\
   Si:S&  1.197& 0.125& 1.545& 0.092\\
    S:O&  0.212& 2.990& 0.000& 0.137\\
   Ne:O&  0.347& 2.590& 0.000& 0.120\\
  \enddata
  \label{tbl:tinsens}
  \tablecomments{The coefficients are to be used as defined by
    Equation~\ref{eq:tinsensfinal}. The last column, $\sigma$, is the
    standard deviation of the emissivity ratio, which is never
    perfectly flat, and coefficient $a_0$ represents the mean of the
    ratio.} 
\end{deluxetable}

 \begin{deluxetable}{ccc}
  \tablecolumns{3}
  \tablewidth{4in}
  \tablecaption{Abundance Ratios\tablenotemark{a}}
  \tablehead{
    \colhead{Elements}&
    \colhead{$r(T_{insens})$}&
    \colhead{$r(EM)$}
  }
  \startdata
  Ne:Mg&  1.7    (0.3)&      2.2   (0.2)\\
  Mg:Si&  1.0    (0.1)&      1.1   (0.1)\\
   Si:S&  1.2    (0.2)&      1.4   (0.4)\\
    S:O&  1.4    (1.0)&      1.9   (0.8)\\
   Ne:O&  4.9    (2.8)&      6.5   (2.3)\\
   Mg:O\tablenotemark{b}&  2.9    (1.8)&      2.9   (1.0)\\
   Si:O\tablenotemark{b}&  1.8    (0.9)&      2.6   (0.9)\\
  \enddata
  \label{tbl:abundratios}
  \tablecomments{$r(T_{insens})$ gives the abundance ratios from the
    temperature-insensitive method, while $r(EM)$ gives ratios from
    the emission measure reconstruction (see Table~\ref{tbl:abund}).
    Values in parentheses are $1\sigma$ statistical uncertainties.}
  \tablenotetext{a}{The ratios are of relative Solar photospheric abundances.}
  \tablenotetext{b}{Derived from preceding ratios - not determined directly.}
\end{deluxetable}


\clearpage
\section{Discussion, Interpretation}

The recent determination by \citet{Herbig:Griffin:2006} that \tooe is
a moderate mass pre-main-sequence spectroscopic binary is very
important in the context of stellar evolution and X-ray activity.
When \tooe\ arrives on the main sequence, we expect it to be faint or
non-detectable in soft X-rays.  Yet at the age of 0.5 Myr, it is the
second-brightest steady X-ray source in the Orion Trapezium.  The
binary system has $L_x = 1.2 \times 10^{32}\, \mathrm{ergs\,s\mone}$,
and given an optical luminosity of $29 \Lsun$
\citep{Herbig:Griffin:2006} it thus has $L_x/L_\mathrm{bol} =
10\mthree$ for the pair.  This value is near the saturation limit of
coronally active stars \citep{Prosser:Randich:al:1996}.  The X-ray
emission is similar to other magnetically active stars, having a broad
temperature distribution and narrow emission lines.  Hence we surmise
that \tooe has dynamo activity and probably has strong convection
zones.  
\Vtworev 
This is also in accordance with evolutionary models of stellar
interiors \citep{Siess:al:2000} which indicate a substantial
convection zone for stars like the \tooe\ components \citep[also see
Figure 1 of][]{Stelzer:al:2005}.  

\Vtworev 
\Vtworev 
Prior to the \citet{Herbig:Griffin:2006} determination that \tooe is a
binary of G-type stars, \tooe was considered to be a B5-star.
\citet{Schulz:2003} considered the X-rays to be a hybrid of wind shock
emission and magnetically confined winds, but they did note a striking
similarity to active coronal sources.  \citet{Stelzer:al:2005}
interpreted the emission as from a weak wind, but noted unusual
emission characteristics, such as due to an extended magnetosphere and
magnetically confined wind shocks.  It is now clear in hindsight,
given the spectral types, that emission is coronal in nature.

\torie did not show any distinct flares during our observations, which
is somewhat unusual for coronally active stars. This lack of activity
is consistent, however, with the long intervals of constant flux seen
by \citet{Stelzer:al:2005}.  There was one observation in which the
flux was {\em lower} than our average (see Figure~\ref{fig:fluxlc}),
and examination of the spectra shows that this is manifested in
diminished short wavelength flux (below about 10\AA).  This variation
between observations cannot be attributed to rotational phase
dependence since it does not repeat.  We also found no significant
variability within any of our observations.  It is possible that the
system is so coronally active that a significant proportion of the
average flux is from continuously visible flares, which would also
give rise to the dominant emission measure peak at $\log T = 7.3$, but
one observation had a bit less flaring.

The \hetgs flux was about half that reported by
\citet{Stelzer:al:2005} from heavily piled, low resolution spectra.
The \hetgs flux calibration is accurate to about 5\%.  Since flux for
our observations was effectively constant and spans the time of the
COUP observations, the difference in flux without obvious flares is
unusual.  Since the \citet{Stelzer:al:2005} analysis was made
difficult by the high photon pile-up in the core of \torie, they
resorted to spectral analysis of photons only from the wings of the
point-spread-function.  We reanalyzed spectra for \torie from one of
the COUP datasets, observation ID 4373, for which the flux was
constant.  We used an extraction radius of 2.25 arcsec centered on the
source, including piled photons and made standard responses for the
region.  To fit the spectrum, we used the pileup model of
\citet{DavisJE:2001} as implemented in ISIS. Pileup is a very
non-linear process; there can be multiple solutions since the
count-rate first saturates with increasing fluence, then can decrease
as events are rejected from telemetry.  Finally, for extremely high
pileup, the count rate can again increase when the core is fully
saturated and the wings grow.  We used two-temperature component,
absorbed \aped plasma models, similar to those of
\citet{Stelzer:al:2005}, and used Monte-Carlo techniques to explore
parameter space, given the multi-valued nature of pileup fitting and
the possibly degenerate nature of the models.  While we found
solutions with fluxes similar to those presented by
\citet{Stelzer:al:2005}, we found equally acceptable solutions
(reduced $\chi^2 < 1.3$ ) with fluxes comparable to our \hetgs-derived
values. Incidentally, all our fits to this one spectrum preferred
$N_H\gtrsim3\times10^{21}\mathrm{cm\mtwo}$, a bit larger than our
adopted value of $2\times10^{21}\mathrm{cm\mtwo}$.  We conclude that the
\citet{Stelzer:al:2005} flux is probably in error and the source is
probably steady outside of distinct flares with a flux of about
$3\times10^{-12}\,\mathrm{ergs\,cm\mtwo s\mone}$.

\Vtworev 
The radial velocities determined from lines are consistent with the
orbital dynamics.  Given a peak-to-peak orbital radial velocity
amplitude of $160\,\mathrm{km\,s\mone}$, we have marginal sensitivity
for detection of the orbital modulation if emission were dominated by
one stellar component (see the confidence limits in
Table~\ref{tbl:log} and Figure~\ref{fig:fluxlc}).  Variability over
the time period of observations could destroy any orbital systematic
radial velocities in X-ray lines if the relative activity level of the
two stellar components changed.  Our phase coverage is also poor, but
the lack of significant velocity offsets at phases of maximum orbital
velocity separation suggests that both stellar components are roughly
equal in X-ray emission.  The marginal detection of line broadening,
particularly at these same phases, is consistent with the broadening
being due to orbital velocity effects.  We conclude that the lines are
similar to other coronal sources - narrow, and effectively unresolved.

\Vtworev 
The absolute abundances are rather low when compared to other coronal
sources.  If we compare \tooe to the abundances derived from
low-resolution COUP spectra of \citet{Maggio:Flaccomio:al:2007}, they
are not only lower by a factor of 5 or more in general, the ratios are
also different --- they are, in fact, uncorrelated.  This probably has
as much to do with different methods and spectral resolutions than
with intrinsic differences between \tooe and average Orion stars.
\citet{Maggio:Flaccomio:al:2007} use two temperature component fits,
and these cannot accurately reproduce abundances and emission measures
for realistic, continuous emission measure distribution plasmas.  If a
fitted temperature component is off the peak of some ion's temperature
of peak emissivity, and there is actually plasma at that temperature,
then a two-temperature model will artificially increase the abundance
of that element in order to reproduce the flux. Comparison of low and
high resolution results with different modeling approach is in general
not meaningful.

\subsection{Loop Sizes}

\note{update loop discussion - intro}
The geometric structure of stellar coronae is largely an open
question, and is relevant to energetics, variability, and likelihood of
interactions with stellar companions or disks.  While there are many
uncertain parameters, we can provide order-of-magnitude estimates via
several methods to show that loops could be compact (small fraction of
the stellar radii of $7 \rsun$; see Table~\ref{tbl:starinfo}), or
comparable to the stellar radii and thus a significant fraction of
the stellar separation (the semi-major axis is about $ 2.5 R_*$; see
\citet{Herbig:Griffin:2006}).

If we assume that the X-ray emission originates in an ensemble of
identical semi-circular loops, we can estimate the order of magnitude
of the loops' radius.  The loop radius (or height if vertically
oriented) relative to the stellar radius can be expressed as
\begin{equation}
  \label{eq:nlops}
h = E_{51}^{1/3}\, R_{10}^{-1}\, n_{10}^{-2/3}\, N_{2}^{-1/3} \alpha_{-1}^{-2/3}  
\end{equation}
in which $E$ is the volume emission measure, $R$ the stellar radius,
$N$ the number of identical loops, $n$ the electron density, and
$\alpha$ is the loop aspect ratio (cross sectional radius to height,
$\le1$), and the subscripts indicate the power of 10 scale factor (all
$cgs$ or unit-less quantities).  Only two of these parameters are well
determined, $E = 1.2 \times 10^{55}\, \mathrm{cm\mthree}$ from our
X-ray spectral modeling, and $R \sim 5\times 10^{11} \, \mathrm{cm}$
from the radial velocity curve analysis of
\citet{Herbig:Griffin:2006}.  Densities are poorly constrained; we
will adopt $10^{12} \, \mathrm{cm\mthree}$ for argument (see
Figure~\ref{fig:triplets}).  From Solar loops, $\alpha$ is about 0.1,
and given the lack of variability in \tooe (to about $10\%$ accuracy;
see Figure~\ref{fig:fluxlc}), we will let $N=100$.  If we assume that
the emission is divided equally between the two binary stellar
components, then with these parameters we obtain $h \sim 0.02$,
implying that the coronae are compact.  For a density of
$10^{10}\,\mathrm{cm\mthree}$, this height increases by a factor of 20
to about $0.4$, a significant fraction of the orbital separation.
%

We can also estimate loop parameters using hydrodynamical models from
the flare temporal and spectral properties.  Assuming that a single
flaring loop dominates the emission, \citet{Reale:2007} expressed the
loop size (his Equation 12) as
$L_9 \sim 3 (T_0/T_M)^2\,T_{0,7}^{1/2}\, \tau_{M,3}$. 
Here $L_9$ is the loop half-length in units of $10^9$ cm, $T_0$ is the
maximum temperature during the flare (and $T_{0,7}$ is the same in
units of $10^7$ K), $T_M$ is the temperature at which maximum density
occurs, and $\tau_{M,3}$ is the time from flare start (in ks) at which
maximum density occurs.  To apply such a model in detail, we would
need the evolution of emission measure (a proxy for density) and
temperature (from time-resolved spectra) from the rise to the decay of
a flare.  We do not have such, so we will make some reasonable
approximations to obtain an order-of-magnitude hydrodynamical loop
size.

The COUP observation of \tooe\ detected a flare
\citep{Stelzer:al:2005}; from this, we estimate that $\tau_{m,3} =
25$.  We have an emission measure distribution; we assume that the
hotter peak and hot tail represents the integrated history over many
flares.  From time-resolved analyses of other stars, we have seen that
such a hot peak can be directly attributed to flaring
\citep{Huenemoerder:01, Gudel:Audard:al:2004}.  We will thus assume
that the flare mean temperature of maximum density (or maximum
emission measure) corresponds to our strongest peak, or $\log T_M =
7.3$ (see Figure~\ref{fig:dem}).  We will assume that the hot tail
of the $EMD$ represents the maximum flare temperature.  This is less
well defined, and we adopt $\log T_0 = 7.8$, which is where there is
an inflection in our $EMD$.  From these parameters we find a relative
loop half-length of about $3.8$ stellar radii, or a height (for
semi-circular vertical loops) of $2.4$ stellar radii.

We can also adopt flare parameters from other giant stars, such as
HR~9024 \citep{Testa:al:2007a} (also see Table~\ref{tbl:starinfo}), for
which $\log T_0 = 7.9$ and $T_0/T_M = 1.4$.  Thus, if \tooe flare
temperatures and densities are similar to those on HR~9024, the loop
half-length is about 1 stellar radius (or a height of 0.6 stellar
radii).

Since these are order-of-magnitude estimates, we conclude that flare
loops can be of order the stellar radius.  In sum, an origin of the
emission from magnetically confined coronal loops is not unreasonable.
To better determine the coronal geometry, more information is needed,
such as more stringent constraints on density, detection of rotational
modulation, or time-resolved spectroscopy of a large flare.

\section{Comparison to Other Stars}\label{sec:comparison}

\begin{figure}[htb]
  \includegraphics[angle=0,scale=1.0]{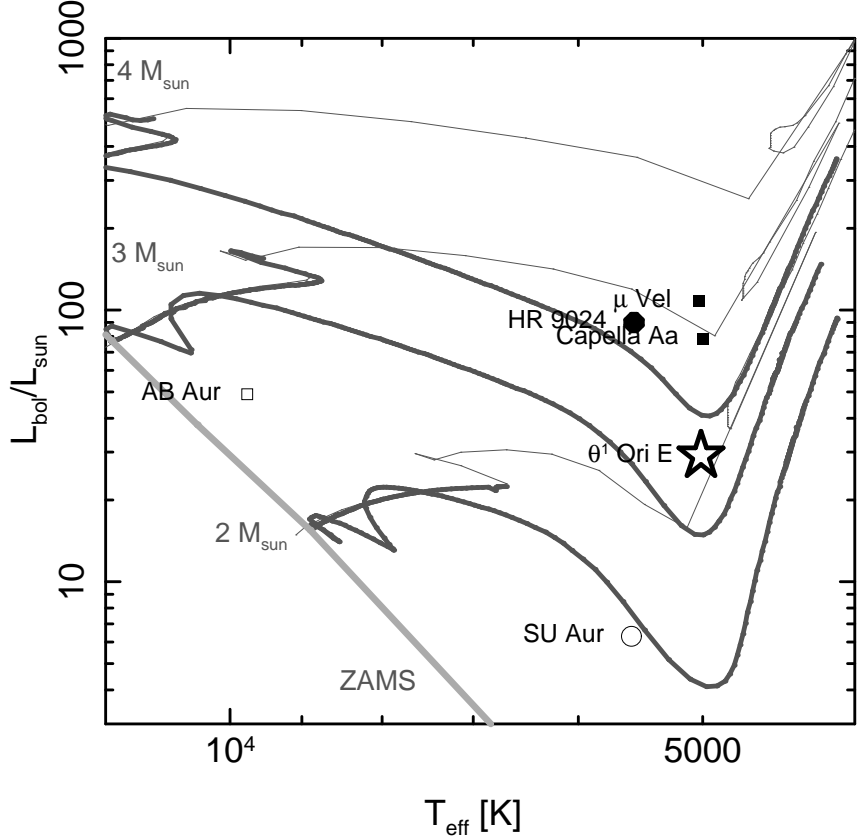} 
  \caption{The stars listed in Table~\ref{tbl:starinfo} are shown on a
    Temperature-Luminosity diagram. Pre-main-sequence stars are shown
    with open symbols and post-main-sequence stars are shown as solid
    symbols.  The shapes encode $L_x/L_\mathrm{bol}$ as a ``star'' for
    $\ge10^{-3}$, circle for $\ge 10^{-4}$ to $<10^{-3}$, and a square
    for $<10^{-5}$ (see Table~\ref{tbl:starinfo} for values).  The
    pre-main-sequence evolutionary tracks (thick gray lines) and
    zero-age main-sequence (ZAMS) are from \citet{Siess:al:2000}.  The
    post-main-sequence tracks (thin gray lines) are from
    \citet{Schaller:al:1992}.  The tracks are shown for 2, 3, and 4
    \Msun.  The post-main-sequence tracks were scaled slightly (0--3\%
    in temperature, 8--12\% in luminosity) to better coincide with the
    pre-main-sequence tracks.) 
 }
  \label{fig:starshrd}
\end{figure}

To understand the nature of the X-ray emission from \tooe, we must
examine it in the evolutionary context of stars of similar mass.  We
have collected information for several other stars, both
pre-main-sequence and post-main-sequence, with masses ranging from 2
to $3.5 \msun$. Information and sources are listed in
Table~\ref{tbl:starinfo} and in Figure~\ref{fig:starshrd} we show the
objects on a temperature-luminosity diagram along with evolutionary
tracks.  Of the sample, \tooe has the highest relative X-ray luminosity,
being as high in $L_x / L_{bol}$ as ``saturated'' short period active binaries
\citep{Vilhu:Rucinski:1983,Cruddace:Dupree:1984,Prosser:Randich:al:1996},
even though its period is somewhat longer than those systems.  If we
look to the future of \tooe's evolution and consider AB~Aur, we see
that $L_x/L_\mathrm{bol}$ may become much smaller; we expect main-sequence
\Vtworev 
A-type stars to be very faint or undetected in X-rays (for reference,
a spectral type A0~V star has a mass of $\sim 2\msun$, and a B5~V of $\sim6
\msun$).

\begin{figure}[htb]
  \includegraphics[angle=0,scale=1.0]{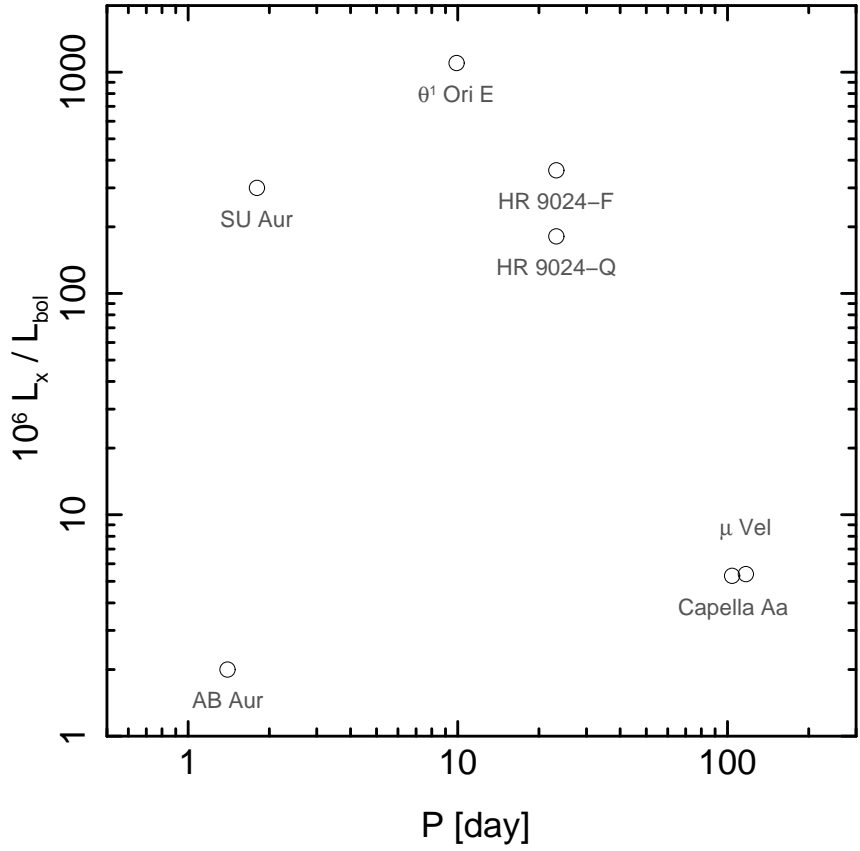} 
  \caption{The normalized X-ray luminosity for stars listed in
    Table~\ref{tbl:starinfo} are shown as a function of their
    periods.  The stars largely follow an expected rotation-activity
    trend, except for AB Aur, which is an A-star near the
    main-sequence and is expected to be faint in X-rays.
    $L_x/L_\mathrm{bol}$ for \tooe\ is for the system; if each stellar
    component contributes equally, its value should be divided by two.
  }
  \label{fig:starspd}
\end{figure}

\Vtworev 
Stellar rotation is well known to be a key factor in magnetic dynamo
generation.  If we compare the sample's X-ray activity as a function
of period (Figure~\ref{fig:starspd}), excluding AB Aur, we see a
strong anti-correlation which holds for both pre- and
post-main-sequence objects. This is similar to the behavior of active
giants, binaries, or main sequence late-type stellar coronae
\citep{Walter:81, Pizzolato:Maggio:al:2003, Gondoin:2005}.  AB~Aur, in
spite of its short period, has very low activity; it is approaching
the low-activity main-sequence era of its life, and may have a
radically different emission mechanism, such as from wind or accretion
affects, with lower $L_x$ and X-ray temperatures
\citep{Telleschi:Gudel:al:2007}.  It's convective zone, necessary for
magnetic dynamo generation, is quite small, being less than 1\% of the
stellar radius, compared to about 20\% for the other stars
\citep[][their Figure 1]{Siess:al:2000,Stelzer:al:2005}, so it is
reasonable to exclude it from period-activity relations of stars with
significant convective regions.
\Vtworev 

\tooe has a very hot corona, characterized by an emission measure
distribution with a strong peak at about 20 MK (see
Figure~\ref{fig:dem}).  It is similar to SU~Aur and HR~9024 (see
Table~\ref{tbl:starinfo}), two objects which displayed strong flares
during their X-ray observations.  A hot emission measure peak has been
directly identified with flares \citep{Huenemoerder:01,
  Gudel:Audard:al:2004}.  In this context, it is curious that during
all the COUP and HETG exposures, only one distinct flare was seen
\citep{Stelzer:al:2005}.  We can only speculate that perhaps \tooe is
so active that the flare rate is so high that they are nearly always
superimposed and create a nearly constant flux. Such was found
plausible for Orion's low-mass stars by \citet{Caramazza:al:2007}.
Our single low-flux \hetgs observation (see Figure~\ref{fig:fluxlc})
did have a relative deficit in short wavelength flux
($<5\,\mathrm{\AA}$), a region sensitive to the highest temperature
plasmas; it is consistent with diminished flare activity.  To sustain
continuous flaring, there has to be a continuous source of erupting
magnetic fields and their reconnection.  The pressure scale-height for
a hot, low-density plasma is a significant fraction of the binary
stellar separation of \tooe.  There could be star-to-star magnetic
reconnections sustained at a fairly high level by the binary proximity
and dynamo action generating sufficiently large loops.

The presence of a very hot corona and the low probability of distinct
flares is common to several evolved giants, such as HR~9024,
$\mu$~Vel, 31~Com, or IM~Peg \citep{Testa:al:2007a,
  Testa:Drake:al:2004b, Ayres:Hodges:al:2007}.  While this is not
understood, it could be a significant trait of the coronal heating
mechanisms.

Another distinguishing characteristic of \tooe is the relatively low
mean metal abundance.  Table~\ref{tbl:abund} shows relative elemental
abundances from the $EMD$ analysis. All abundances are significantly
below unity with oxygen at an extremely low value.  When compared to
abundances deduced in a similar analysis of \tooc \citep{Schulz:2003},
its massive neighbor within the Orion Trapezium, then there are a few
remarkable differences to note. Values for Ne, Al, and Ca seem very
similar within uncertainties, while Mg, Si, S, and Ar are very
different, being near or above unity in \tooc; values for O and Fe are
even lower than in \tooc.  Figure~\ref{fig:abund} also compares
coronal with average Orion stellar photospheric and nebular values.
Since the photospheric and nebular abundances are all near unity, it
seems that abundances from the hot X-ray plasmas are fundamentally
different.  A two-temperature modeling of \tooc could reconcile
deficient O and Ne values by requiring a significantly higher column
density \citep{gagne2005}, as would forcing the Ne/O ratio to be
similar to other coronae \citep{Drake:Testa:2005}, though Fe would
still remain low.  In the case of \tooe, the application of a higher
column will not change the low values for Mg and higher-Z elements,
but also not enough for O, which at $A(O)=0.06$ is extremely low. We
also find that multi-temperature plasmas more accurately describe the
spectra in both stars and thus see trends in Orion Trapezium stars
which include some neon deficiency but clearly hot plasmas with low
\note{updated}
iron and oxygen abundances.  For pre-main-sequence stars, depletion of
metals has been explained by formation of dust in the disk, and the
remaining gas, seen heated to X-ray temperatures in an accretion
shock, is the metal-poor material \citep{Drake:al:2005}.  Magnetic
coronae generally show low metals, but tend to have higher Ne and O
(e.g., II Peg, \citet{Huenemoerder:01}, or HR~1099, \citet{Drake:01}).
Hot plasmas of Orion's stars, if the two studied are representative, seem
to be different.  There is no theoretical explanation for coronal
abundances, but these differences may be clues to fractionation
mechanisms.

 \begin{deluxetable}{rcccccc}
  \tablecolumns{7}
  \tablewidth{0in}
  \tablecaption{Stellar Comparison}
  \tablehead{
    \colhead{Quantity}&
    \colhead{\tooe\tablenotemark{a}}&
    \colhead{SU Aur}&
    \colhead{AB Aur}&
    \colhead{$\mu$ Vel}&
    \colhead{HR~9024\tablenotemark{b}}&
    \colhead{Capella Aa\tablenotemark{c}}
  }
  \startdata
  Spec.Type.  & G8 III+G8 III&G2 IV&  A0   &   G5 III&    G1 III&  G8 III\\
  $T_\mathrm{eff}$ [K]& 5012         &5550 &  9750 &  5030   & 5530     &  5000  \\
  Age [My]    & 0.5          &4    &  4 &     360&   post-ms&  525   \\
  $M/$\Msun   & 3.5          &2.0  &  2.7  &    3    &      2.9 &    2.7 \\
  $R/$\Rsun   & 7            &2.6  &  2.3  &  14.4   &   13.6   &  12.2  \\
  $g/g_\odot$  & 0.07         &0.3  &  0.5  &  0.01   &   0.02   &  0.02  \\
  $P$ [day]   & 9.9          &1.8  &  1.4: &   117     &  23.2    &  104   \\
  $v\sin i\, \mathrm{[km s\mone]}$
              & 37           &66   &  80   &   6.2   &  22      &   5    \\
  $L_\mathrm{bol}/$\lsun
              & 29           &6.3  &  49        &  108    &  90      & 78.5   \\
  $L_x/10^{30}\,\mathrm{[ergs\, s\mone]}$
              & 121          &8    &  0.4       &  2.2    &  63,125  &   1.6  \\
  $(L_x/L_\mathrm{bol})/10^{-6}$
              & 1100         &300  &  2    &    5.4  & 181,360  &  5.3   \\
  $VEM/10^{52}\,\mathrm{[cm\mthree]}$
              & 1200         &50  &  5    &    8    &  200,800 &   4    \\
  $\overline{T}_x$ [MK]\tablenotemark{d}& 20           &20   &  4.7  &  7.9    &  30,80   &   6.3  \\
  $A(Fe)$\tablenotemark{e}
              & 0.1          &0.6  &  0.3  &   1.8   &  0.2,0.7 &   0.8  \\
  $A(O)$\tablenotemark{e}
              & 0.1          &0.3  &  0.2  &   0.7   &  0.6,1.1 &   0.4  \\
  $A(Ne)$\tablenotemark{e}
              & 0.4          &1.3  &  0.6  &  1.4    &  1.1,1.2 &   0.7  \\
  \enddata
  \label{tbl:starinfo}
  \tablenotetext{a}{X-ray properties are for the binary system; if
    each stellar component contributes equally, values should be
    divided by two.}  
  \tablenotetext{b}{When two quantities are given
    they are for quiescent and flare states, respectively.}
  \tablenotetext{c}{X-ray properties assume that component Aa
    dominates.}  \tablenotetext{d}{A qualitative temperature of the
    high energy emission measure distribution, adopted from visual
    inspection of published curves or few-$T$ fits.}
  \tablenotetext{e}{Abundances are relative to Solar.}
  \tablecomments{Sources:
    {\bf $\mathbf\theta^1$ Ori~E:}    \citet{Herbig:Griffin:2006};  this paper.
    {\bf SU Aur:}  \citet{Franciosini:Scelsi:al:2007, Robrade:Schmitt:2006, DeWarf:Sepinsky:al:2003}.
    {\bf AB Aur:}  \citet{Telleschi:Gudel:al:2007}.
    {\bf $\mathbf\mu$ Vel:}  \citet{Ayres:Hodges:al:2007, Wood:Redfield:al:2005}.
    {\bf HR~9024:} \citet{Ayres:Hodges:al:2007, Testa:al:2007a}.
    {\bf Capella:}  \citet{Ishibashi:al:2006, Canizares:00,  Hummel:Armstrong:al:1994, 
      Ness:Brickhouse:al:03, Gu:Gupta:al:2006}.
  }
\end{deluxetable}
\clearpage 

\clearpage

\Vtworev
\clearpage 

%


\section{Conclusions}

\torie is perhaps the only case known of a pre-main-sequence 3 Solar
mass G-star binary.  As such, it holds an important place in our
understanding of X-ray dynamo generation and evolution.  We believe
that when it reaches the main sequence, it will emit a negligible
fraction of its luminosity in X-rays.  Yet now it is the second
brightest X-ray source in the Trapezium.  Furthermore, its relative
X-ray luminosity ($L_x/L_\mathrm{bol}$) makes it as strong as any of
the coronally active binaries.  Thus we conclude that as moderate mass
stars collapse toward the main sequence, they go through a phase of
strong magnetic dynamo generation, very similar or identical to that
of coronally active late-type stars which sustain a convective zone
and shear-generated magnetic dynamo.
\Vtworev 
Since A-stars are dark or at most quite faint in X-rays
\citep{Schroeder:Schmitt:2007}, at some point, the dynamo vanishes,
and the magnetic fields dissipate.
\Vtworev 
AB Aur, which is near the main sequence and relatively faint in
X-rays, is a possible future state of \tooe.
It is also clear from the post-main-sequence objects that a dynamo can
be generated as stars of this mass evolve into giants.

\note{updated}
Hot plasma abundances of \tooe (as well as of \tooc) are different
from coronally active stars.  \tooe is hot and has fairly steady X-ray
emission.  The high temperature is either due to unresolved flares, or
to an unknown mechanism which also may be common to other active
G-giants.  The temperature structure, abundances, and low variability
may be clues to plasma heating mechanisms.  Continuing high-resolution
spectroscopic studies of Orion stars will show us if some of these
patterns are common in the Orion Nebular Cluster.


\acknowledgments

\paragraph{Facilities:} \facility{CXO (HETGS)}

\paragraph{Acknowledgments} Support for this work was provided by the
National Aeronautics and Space Administration through the Smithsonian
Astrophysical Observatory contract SV3-73016 to MIT for Support of the
Chandra X-Ray Center, which is operated by the Smithsonian
Astrophysical Observatory for and on behalf of the National
Aeronautics Space Administration under contract NAS8-03060.

%
 


\end{document}